\documentclass[preprint]{jpsj2}
\title{Effect of in-plane line defects on field-tuned 
superconductor-insulator transition behavior in homogeneous thin film}

\author{
Kiyokazu Myojin and Ryusuke Ikeda
}

\inst{
Department of Physics, Kyoto University, Kyoto 606-8502
}

\recdate{\today}

\abst{Field-tuned superconductor-insulator transition (FSIT) behavior in 2D isotropic and homogeneous thin films is usually accompanied by a nonvanishing critical resistance at low $T$. It is shown that, in a 2D film including line defects paralle to each other but with random positions perpendicular to them, the (apparent) critical resistance in low $T$ limit vanishes, as in the 1D quantum superconducting (SC) transition, under a current parallel to the line defects. This 1D-like critical resistive behavior is more clearly seen in systems with weaker point disorder and may be useful in clarifying whether the true origin of FSIT behavior in the parent superconductor is the glass fluctuation or the quantum SC fluctuation. As a by-product of the present calculation, it is also pointed out that, in 2D films with line-like defects with a long but {\it finite} correlation length parallel to the lines, a quantum metallic behavior intervening the insulating and SC ones appears in the resistivity curves. 
}

\kword{
Quantum Fluctuation, Vortex States, Superconductor-Insulator Transition
}

\begin{document}
\maketitle

\section{Introduction}

 In homogeneously disordered thin superconducting (SC) films, the resistivity shows an insulating behavior in higher fields, while it decreases upon cooling in lower fields as a precursor of a SC transition. At a critical field separating these two regimes, the resistivity approaches a finite value upon cooling. It is well understood that this field-tuned superconductor-insulator (FSIT) resistive behavior in two-dimensional (2D) impure superconductors is induced by a quantum fluctuation enhanced at lower temperatures. However, two different origins leading to the FSIT behavior have been proposed so far \cite{MPAF,IR,RI96,Gant}. One is the quantum {\it critical} fluctuation accompanying the continuous melting transition of the 2D vortex-glass (VG) state at $T=0$. The other is a consequence of the thermal to quantum crossover behavior in the conventional Aslamasov-Larkin fluctuation conductivity $\sigma_{s, \, jj}$ ($j=x$, $y$): The thermal fluctuation enhances $\sigma_{s, \, jj}$ upon cooling, while the quantum fluctuation is insulating \cite{RI96} and reduces $\sigma_{s, \, jj}$ upon cooling. Since, as the magnetic field is lowered, the fluctuation is weakened, and the quantum fluctuation is changed into the thermal one, just the SC fluctuation itself may induce the FSIT behavior. However, this picture, first argued in Ref.3, was found not to be justified once the vortex pinning effect is taken into account at the microscopic level for the homogeneously disordered SC films \cite{IAR}. Nevertheless, it has been argued recently that the FSIT behavior in NdCuO is a consequence of the conventional fluctuation conductivities including $\sigma_{s, \, jj}$ \cite{Gant}. Under such situation, it will be valuable to propose an experiment for judging which of the two origins of an FSIT behavior plays a dominant role.

\begin{figure}
\scalebox{1.7}[1.7]{\includegraphics{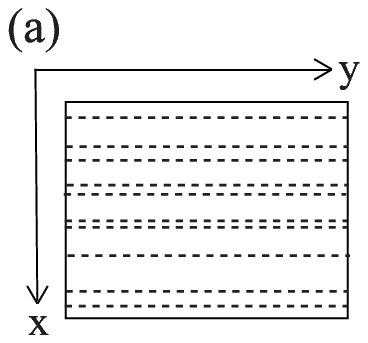}}
\hspace{10mm}
\scalebox{1.7}[1.7]{\includegraphics{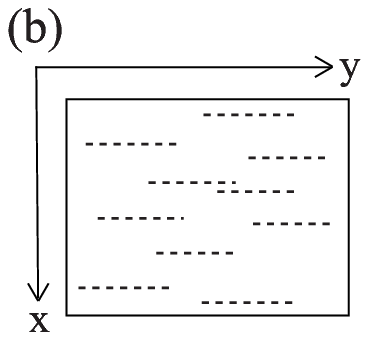}}
\caption{(a) Sketch of a SC thin film under a magnetic field parallel to ${\hat z}$ and with line defects (dashed curves) parallel to the $y$ axis. (b) The corresponding figure including line-like defects with a finite correlation length along the $y$ axis in place of the line defects in (a). 
}
\end{figure}

In this paper, effects of line defects on resistivity curves in a homogeneously disordered thin SC film are considered at relatively low temperatures. The film sample is assumed to be isotropic prior to introducing the line defects running along with one another, and their positions perpendicular to their extended direction are not periodic but random. As sketched in Fig.1(a), such line defects may be artificially introduced in real amorphous thin films. It is found by examining the SC part of conductivity that the line defects make the scaling behavior of resistivity anisotropic : The line defects merely affect the magnitude of the SC part of conductivity when the current is perpendicular to the line defects, while they make the nonvanishing conductance \cite{MPAF,IR} at the (apparent) critical field $H_c$ divergent for a current parallel to the line defects. We find through numerical analysis of resistivity that this anisotropic behavior appears more remarkably in the case with a smaller $R_n$ (i.e., weaker point disorder), where the flat resistivity curve at a critical field for the perpendicular current is vague reflecting that the FSIT behavior originates largely from the quantum SC fluctuation, while the vanishing critical resistance for the parallel current is well defined. Thus, the present result is relevant to understanding the origin of the FSIT behavior in real systems correctly. As a byproduct, we also examine the case such as Fig.1(b) including line-like defects with a {\it finite} correlation length and find that, reflecting a 1D to 2D crossover upon cooling, the resistivity curves in currents parallel to the line-like defects, close to the critical field, decrease rapidly at intermediate temperatures, while they become flat at the lowest temperatures even in systems with weak quantum fluctuation. Implication of such a nonmonotonic behavior will also be discussed in relation to real observations. \cite{Mason,Vicente,Hebard,Wu}

In sec.II, analytical expressions of conductivities are derived on the basis of the Ginzburg-Landau (GL) model. In sec.III, numerical results of the conductivities in sec.II are shown, and their contents are discussed. Consequences of the present results are discussed in sec.IV.

\section{Model and analytical calculations}

To explain the analysis used in this paper, let us first start with the 2D Ginzburg-Landau action $S_{\rm GL}$ to be derived microscopically 
\begin{eqnarray}
S_{\rm GL} &=& s\int\,d^2 r\biggl\{\,\beta\sum_{\omega}\psi^{\ast}_{\omega}({\bf r})\gamma({\bf Q}^2)|\omega|\psi_{\omega}({\bf r})+\int^{\beta}_{0}\,d\tau\biggl[\,\psi^{\ast}({\bf r},\,\tau)\mu({\bf Q}^2)\psi({\bf r},\,\tau) \nonumber \\
&+& \frac{b}{2}|\psi({\bf r},\,\tau)|^4+u({\bf r})\rho({\bf r},\,\tau) + f({\bf r})\rho_v({\bf r},\,\tau)\,\biggr]\biggr\},
\end{eqnarray}
where $\rho({\bf r},\,\tau) = |\psi({\bf r},\,\tau)|^2$, $\rho_v({\bf r},\,\tau) = [\nabla\times{\bf J}({\bf r},\,\tau)]_z$, 
${\bf J} = \xi_0^2 (\psi^{\ast} {\bf Q}\,\psi + {\rm c.c.})$, ${\bf Q} = -{\rm i}\nabla + 2 \pi{\bf A}/{\phi_{0}}$ is the gauge-invariant gradient, and random potential terms leading to pinning effect of vortices were assumed to be present. As shown elsewhere, in the absence of line defects lying in the plane, the corresponding replicated action rather than eq.(1) itself is derived microscopically. Here, $\psi({\bf r},\,\tau) = \sum_{\omega}\psi_{\omega}({\bf r})\,e^{-{\rm i}\omega\tau}$,  $b > 0$, $s$ is the film thickness, $\tau$ is an imaginary time, and $\omega$ is a Matsubara frequency. Further, when a small electron-hole asymmetry is neglected, $\gamma$ is real and positive so that the dynamics of the pair-field is purely dissipative. The random potentials will be assumed to satisfy the following Gaussian ensembles 
\begin{eqnarray}
\overline{u({\bf r})u({\bf r^{\prime}})} &=& w({\bf r}-{\bf r^{\prime}}) = {\Delta_p}\delta({\bf r}-{\bf r^{\prime}})+\frac{\Delta_l}{\xi_0}\delta(x-x^{\prime}), \nonumber \\
w({\bf k}) &=& \int d^2 r\,w({\bf r})\,e^{-{\rm i}{\bf k}\cdot{\bf r}} = \Delta_{p}+\frac{2\pi\Delta_{l}}{\xi_0}\delta(k_y), \nonumber \\
\overline{f({\bf r})f({\bf r^{\prime}})} &=& w_{\Phi}({\bf r}-{\bf r^{\prime}}), \nonumber \\
w_{\Phi}({\bf k}) &=& \int d^2r\,w_{\Phi}({\bf r})\,e^{-{\rm i}{\bf k}\cdot{\bf r}} = \Delta_p^{(\Phi)}+\frac{2\pi \Delta_l^{(\Phi)}}{\xi_0}\delta(k_y), 
\end{eqnarray}
where $\xi_0$ is the coherence length defined in the ordinary GL region (i.e., in low fields and near $T_c$). Hereafter, the length scale and the pair-field will be treated in dimensionless forms under the scale transformations
${\bf r}/r_{\small H} \to {\bf r}$, and $\psi \, (\beta s)^{1/2} \to \psi$, where $r_{\small H}=\sqrt{\phi_0/(2\pi H)}$.

After performing the random average, we encounter a replicated action $S^{\,n}$ corresponding to eq.(1), where 
\begin{eqnarray}
S^n = S_0^n+S_p^n,
\end{eqnarray}
and the free energy is given by 
\begin{equation}
{\cal F}= \lim_{n \to +0} \frac{Z^n - 1}{n},
\end{equation}
and 
\begin{eqnarray}
{\cal Z}^n = {\rm Tr}_{\psi}\,e^{-S^n}.
\end{eqnarray}
The first term of eq.(3) is the contribution independent of the pinning effect and, using the Landau expansion of the pair-field 
\begin{eqnarray}
\psi_{\omega}^{(\alpha)}({\bf r}) &=& \sum_{n,\,p}\varphi_n^{(\alpha)}(p,\,\omega){\tilde u}_{n,\,p}({\bf r}),
\end{eqnarray}
becomes 
\begin{eqnarray}
S_0^n &=& \sum_{\alpha}\sum_{\omega}\biggl[\,(\mu_{0}+\gamma_ 0|\omega|)\sum_p |\varphi_0^{(\alpha)}(p,\,\omega)|^2+({\cal G}_1 (\omega))^{-1}\sum_{p}|\varphi_1^{(\alpha)}(p,\,\omega)|^2 \nonumber \\
&+& \frac{b}{4\pi r_{\small H}^2 N_v s}\beta^{-1}\sum_{\bf k}{\tilde \rho}_0^{(\alpha)}({\bf k},\,\omega){\tilde \rho}_0^{(\alpha)}(-{\bf k},\,-\omega)\biggr],
\end{eqnarray}
where $\alpha$ is a replica index, 
\begin{eqnarray}
{\tilde \rho}_0^{(\alpha)}({\bf k},\,\omega) = \sum_{p,\,\omega^{\prime}}\,\exp\biggl({{\rm i} p k_x -\frac{k^2}{4}}\biggr) \, (\varphi_0^{(\alpha)}(p_-,\,\omega^{\prime}))^{\ast}\varphi_0^{(\alpha)}(p_+,\,\omega+\omega^{\prime})
\end{eqnarray}
($p_{\pm}=p\pm{k_{y}}/{2}$), and 
\begin{eqnarray}
{\cal G}_1 (\omega) = \bigl\langle\,|\varphi_1 (p,\,\omega)|^2\,\bigr\rangle = (\mu_1+\gamma_1|\omega|)^{-1}.
\end{eqnarray}
Here, ${\tilde u}_{n,\,p}({\bf r})$ is the eigenfunction in the $n$-th LL, and only the lowest ($n=0$) and the next lowest ($n=1$) LLs were kept. Further, $N_v$ is the number of field-induced vortices, and the dependences of $\gamma_n$ and $\mu_n$ on the LL index $n$ stem from the ${\bf Q}$-dependence of the corresponding coefficients in eq.(1).

Further, note that, in eq.(9), $\mu_1$ is the bare mass of the $n=1$ LL fluctuation. That is, we have included the $n=1$ LL fluctuation only at the Gaussian level. Since this fluctuation is heavy, and $\mu_1$ is of order unity in  high fields and low temperatures relevant to the quantum fluctuation phenomena, the renormalization of $n=1$ LL fluctuation due to interactions among fluctuations is negligible. For this reason, the Gaussian approximation for the $n=1$ LL modes is adequate. 
On the other hand, the lowest LL fluctuation should be treated in a fully renormalized form, because the events in equilibrium, such as the vortex lattice melting in clean limit and the glass transitions, occur in the lowest LL in the present high field approximation. For the present purpose of examining resistive behaviors in the vortex liquid region, the one-loop approximation used elsewhere \cite{IAR} for the lowest LL fluctuation propagator ${\cal G}_0(\omega)=\langle\,|\varphi_0(p,\,\omega)|^2\,\rangle$ will be sufficient. Then, ${\cal G}_0(\omega)$ is given by
\begin{eqnarray}
{\cal G}_0 (\omega) &\equiv& \bigl\langle\,|\varphi_0^{(\alpha)}(p,\,\omega)|^2\,\bigr\rangle, \nonumber \\
&=& \biggl(\gamma_0 |\omega|+\mu_0 +\Sigma_0-\frac{{\cal G}_0 (|\omega|)}{2\pi r_{\small H}^2 s}\biggl(\Delta_p +\Delta_l \sqrt{\frac{2\pi}{h}}\biggr)\biggr)^{-1}. \\ 
\Sigma_{0} &=& \frac{b}{2\pi^2 r_{\small H}^2 s\gamma_0}\int_0^{\varepsilon_c}d\varepsilon\frac{\varepsilon}{\varepsilon^2+({\cal G}_0 (0))^{-2}}\coth\biggl(\frac{\beta\varepsilon}{2\gamma_0}\biggr),
\end{eqnarray}
where $h= \xi_0^2/r_{\small H}^2 = 2 \pi H\xi_0^2/\phi_0$. The pinning-induced contribution $S_p^n$ to the action (3) is given by 
\begin{eqnarray}
S_p^n &=& - \, \frac{1}{2}\sum_{\alpha,\,\beta}\int d^{\,2}{\bf r}\int d^{\,2}{\bf r^{\prime}}\biggl[w({\bf r}-{\bf r^{\prime}}){\tilde \rho}^{(\alpha)}({\bf r},\,0){\tilde \rho}^{(\beta)}({\bf r^{\prime}},\,0) 
+ w_{\Phi}({\bf r}-{\bf r^{\prime}}){\tilde \rho}_v^{(\alpha)}({\bf r},\,0){\tilde \rho}_v^{(\beta)}({\bf r^{\prime}},\,0)\biggr] \nonumber \\
&=& - \, \frac{1}{2}\sum_{\alpha,\,\beta}\sum_{\bf k}\biggl[w({\bf k})({\tilde \rho}^{(\alpha)}({\bf k},\,0))^{\ast}{\tilde \rho}^{(\beta)}({\bf k},\,0)+w_{\Phi}({\bf k})({\tilde \rho}_v^{(\alpha)}({\bf k},\,0))^{\ast}{\tilde \rho}_v^{(\beta)}({\bf k},\,0)\biggr].
\end{eqnarray}
where 
\begin{eqnarray}
{\tilde \rho}({\bf k},\,0) &=& \int d^2 r\,{\tilde \rho}({\bf r},\,0)\,e^{-{\rm i}{\bf k}\cdot{\bf r}} \nonumber \\
&=& \sum_{p,\,\omega} (v_{{\bf k}_{\perp}})^{1/2}\,e^{{\rm i} k_x p}\biggl[\varphi_0^{\ast}(p_+,\,\omega)\varphi_0({p_-},\,\omega)-{\rm i}\frac{k_+}{\sqrt{2}}\varphi_1^{\ast}(p_+,\,\omega)\varphi_0(p_-,\,\omega)  \nonumber \\
&-& {\rm i}\frac{k_-}{\sqrt{2}}\varphi_0^{\ast}(p_+,\,\omega)\varphi_1(p_-,\,\omega)\biggr], \\
{\tilde \rho}_v({\bf k},\,0) &=& h  \sum_{p,\,\omega}(v_{\bf k_{\perp}})^{1/2}\,e^{{\rm i}k_x p}\biggl[k^2 \varphi_0^{\ast}(p_+,\,\omega)\varphi_0(p_-,\,\omega) + {\rm i}\frac{k_+}{\sqrt{2}}(k^2-2)\varphi_1^{\ast}(p_+,\,\omega)\varphi_0(p_-,\,\omega) \nonumber \\
&+& {\rm i}\frac{k_-}{\sqrt{2}}(k^2-2)\varphi_0^{\ast}(p_+,\,\omega)\varphi_1(p_{-},\,\omega)\biggr], 
\end{eqnarray}
$k_{\pm} = k_x\pm {\rm i}k_y$, and $k^2 = k_x^2+k_y^2$.

Therefore, $S_p^n$ takes the form
\begin{eqnarray}
S_p^n \!\!\! &=& \!\!\! - \, \frac{1}{2}\sum_{\alpha,\,\beta}\sum_{\bf k}\biggl[\,{\tilde \Delta}_0 ({\bf k}){\tilde \rho}_0^{(\alpha)}({\bf k},\,0){\tilde \rho}_0^{(\beta)}(-{\bf k},\,0) \nonumber \\
&+& \frac{k^2}{2}{\tilde \Delta}_1({\bf k})({\tilde \rho}_1^{(\alpha)}({\bf k},\,0))^{\ast}{\tilde \rho}_1^{(\beta)}({\bf k},\,0)+\frac{k^2}{2}{\tilde \Delta}_1({\bf k}){\tilde \rho}_1^{(\alpha)}(-{\bf k},\,0)({\tilde \rho}_1^{(\beta)}(-{\bf k},\,0))^{\ast} \nonumber \\
&+& \frac{k_+^2}{2}{\tilde \Delta}_1({\bf k}){\tilde \rho}_1^{(\alpha)}(-{\bf k},\,0){\tilde \rho}_1^{(\beta)}({\bf k},\,0)+\frac{k_-^2}{2}{\tilde \Delta}_1({\bf k})({\tilde \rho}_1^{(\alpha)}({\bf k},\,0))^{\ast}({\tilde \rho}_1^{(\beta)}(-{\bf k},\,0))^{\ast} \nonumber \\
&+& {\rm i}\frac{k_+}{\sqrt{2}}{\tilde \Delta}_{01}({\bf k}){\tilde \rho}_0^{(\alpha)}(-{\bf k},\,0){\tilde \rho}_1^{(\beta)}({\bf k},\,0)-{\rm i}\frac{k_-}{\sqrt{2}}{\tilde \Delta}_{01}({\bf k})({\tilde \rho}_1^{(\alpha)}({\bf k},\,0))^{\ast}{\tilde \rho}_0^{(\beta)}({\bf k},\,0) \nonumber \\
&-& {\rm i}\frac{k_+}{\sqrt{2}}{\tilde \Delta}_{01}({\bf k}){\tilde \rho}_1^{(\alpha)}(-{\bf k},\,0){\tilde \rho}_0^{(\beta)}({\bf k},\,0)+{\rm i}\frac{k_-}{\sqrt{2}}{\tilde \Delta}_{01}({\bf k}){\tilde \rho}_0^{(\alpha)}(-{\bf k},\,0)({\tilde \rho}_1^{(\beta)}(-{\bf k},\,0))^{\ast}\biggr],
\end{eqnarray}
where
\begin{eqnarray}
{\tilde \rho}_1^{(\alpha)}({\bf k},\,\omega) &=& \sum_{p,\,\omega^{\prime}}\,\exp\biggl({\rm i}pk_x-\frac{k^2}{4}\biggr)(\varphi_1^{(\alpha)}(p_-,\,\omega^{\prime}))^{\ast}\varphi_0^{(\alpha)}(p_+,\,\omega+\omega^{\prime}), \\ 
{\tilde \Delta}_0({\bf k}) &=& \exp\biggl(-\frac{k^2}{2}\biggr)\biggl[(\Delta_p+h^2 \Delta_p^{(\Phi)}k^4)+\frac{2\pi}{\sqrt{h}}(\Delta_l+h^2 \Delta_l^{(\Phi)}k^4)\delta(k_y)\biggr], \\
{\tilde \Delta}_1({\bf k}) &=& \exp\biggl(-\frac{k^2}{2}\biggr)\biggl[(\Delta_p+h^2 \Delta_p^{(\Phi)}(k^2 -2)^2) \nonumber \\
&+& \frac{2\pi}{\sqrt{h}}(\Delta_l+h^2 \Delta_l^{(\Phi)}(k^2-2)^2)\delta(k_y)\biggr], \\
{\tilde \Delta}_{01}({\bf k}) &=& \exp\biggl(-\frac{k^2}{2}\biggr)\biggl[(\Delta_p+h^2 \Delta_p^{(\Phi)}k^2(k^2 -2)) \nonumber \\
&+& \frac{2\pi}{\sqrt{h}}(\Delta_l+h^2 \Delta_l^{(\Phi)}k^2 (k^2 -2))\delta(k_y)\biggr].
\end{eqnarray}

In the absence of line defects, i.e., when $\Delta_l=\Delta_l^{(\Phi)}=0$, the above expressions can be regarded as being derived microscopically. Describing line defects introduced artifitially from the microscopic standpoint is not easy, and they were incorporated in a phenomenological but conventional manner.

\begin{figure}
\scalebox{0.55}[0.55]{\includegraphics{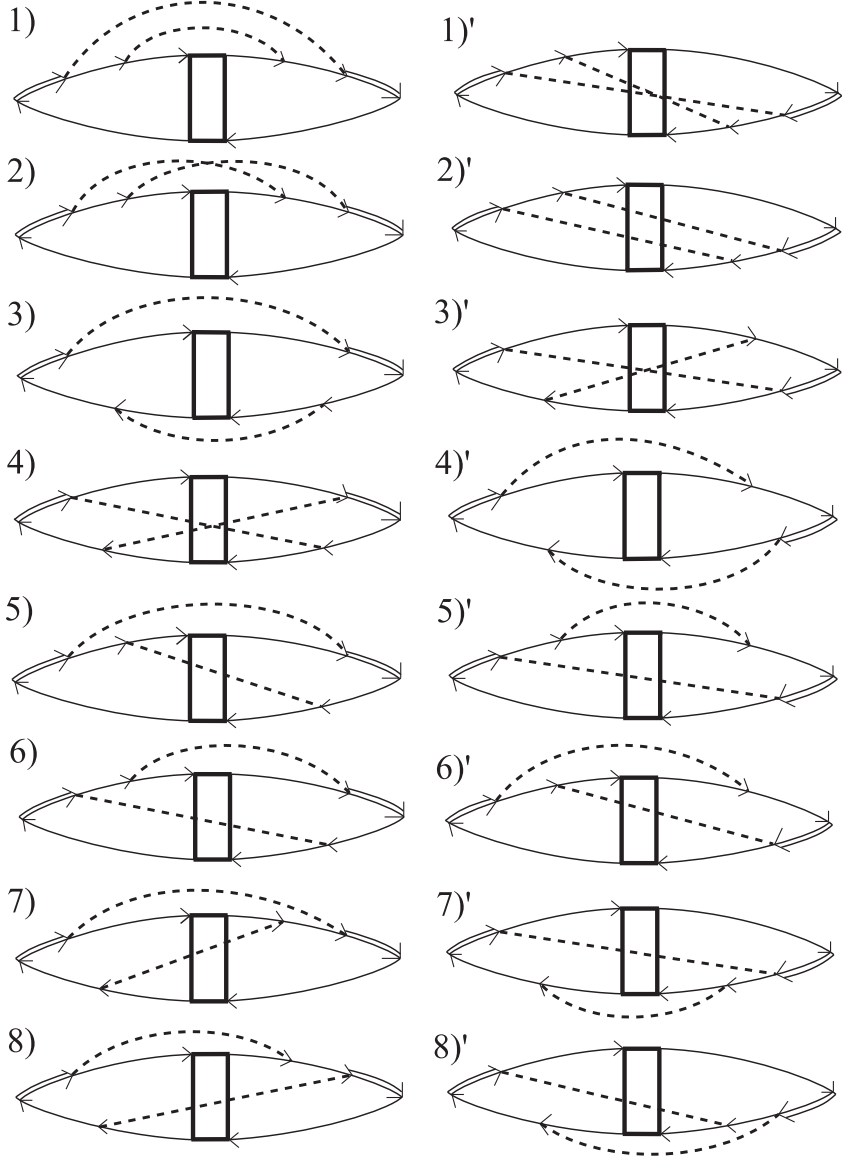}}
\caption{Diagrams representing $\sigma_{{\rm vg}, \, jj}$. Each solid (double-solid) curve denotes ${\cal G}_0(\omega)$ (${\cal G}_1(\omega)$), the dashed curve is the pinning line occurring after the random average, and the open rectangle denotes ${\tilde{\cal G}}_{\rm vg}$. 
}
\end{figure}

We examine hereafter the superconducting part $\sigma_{s, \, jj}$ of conductivity $\sigma_{jj}$ ($j=x$, $y$), according to the Kubo formula, in the form of a current-current correlation function by invoking the high field approximation where the expression of the supercurrent density is linear both in the $n=0$ and $1$ LL modes of the pair-field. In the framework consistent with the one-loop approximation for the lowest LL fluctuation, $\sigma_{s, \, xx}$ takes the form 
\begin{eqnarray}
s \, R_{Q} \, \sigma_{s,\,xx} &=& \biggl(-\frac{\partial}{\partial|\Omega|}\biggr) \frac{\mu_1^2}{N_v \beta} \, {\rm Re} \, \sum_{p,\,p^{\prime},\,\omega}\bigg[\overline{\bigl\langle\varphi_1(p,\,\omega+\Omega)\varphi_1^{\ast}(p^{\prime},\,\omega+\Omega)\bigr\rangle\bigl\langle\varphi_0(p^{\prime},\,\omega)\varphi_0^{\ast}(p,\,\omega)\bigr\rangle} \nonumber \\
&-& \overline{\bigl\langle\varphi_1(p,\,\omega+\Omega)\varphi_0^{\ast}(p^{\prime},\,\omega+\Omega)\bigr\rangle\bigl\langle\varphi_1 (p^{\prime},\,\omega)\varphi_0^{\ast}(p,\,\omega)\bigr\rangle}\bigg]\bigg|_{\Omega\to+0},
\end{eqnarray}
while the corresponding expression of $\sigma_{s, \, yy}$ is given by replacing the minus sign prior to the second term of eq.(20) by a plus sign. Here, $R_{Q}=2\pi\hbar/4e^{\,2}$ is the quantum resistance. Note that, because of line defects introduced along the $y$-axis, the second term of eq.(20) is nonvanishing.

In the presence of pinning disorder, $\sigma_{s,\,\mu \mu}$ can be seen as the sum of the so-called vortex flow term $\sigma_{{\rm fl},\, \mu \mu}$, independent of the vortex-glass fluctuation, and the divergent term $\sigma_{{\rm vg}, \, \mu \mu}$ reflecting a growth of the vortex-glass fluctuation. A method of formulating $\sigma_{{\rm vg}, \, \mu \mu}$ was developed in Ref.10 in the case with only point defects and with no quantum SC fluctuation. As far as the lowest order contribution in $\xi_{\rm vg}^{-1}$ is concerned, where $\xi_{\rm vg}$ is the glass correlation length normalized by $r_{\small H}$, $\sigma_{{\rm vg},\,jj}$ obtained in the lowest order in the pinning strengths consist of the sixteen diagrams in Fig.2. The second term of eq.(20) is given by the sum of diagrams 1)' to 8)'. In the expressions of these diagrams, the vortex-glass correlation function ${\tilde{\cal G}}_{\rm vg}({\bf k}\,;\,\omega,\omega+\Omega)$ inducing the $\xi_{\rm vg}$-dependences is included. In the present case with line defects perpendicular to the field, ${\tilde{\cal G}}_{\rm vg}$ has been studied elsewhere\cite{IM} and, in the so-called ladder approximation, becomes 
\begin{eqnarray}
{\tilde{\cal G}}_{\rm vg}({\bf k}\,; \,\omega,\omega+\Omega) &=& \frac{({\cal G}_0 (0))^2}{r_g+\sum_{\mu = x,\,y}c_{\mu}k_{\mu}^2+r_g^{-1}\gamma_0{\cal G}_0 (0)\,(|\omega|+|\omega+\Omega|)} \nonumber \\
&=& \frac{\displaystyle{\xi_{\rm vg}^2}\,({\cal G}_0 (0))^2}{\displaystyle{1+\xi_{\rm vg}^2\sum_{\mu = x,\,y}c_{\mu}k_{\mu}^2+\xi_{\rm vg}^4 \gamma_0{\cal G}_0 (0)|\omega|+\xi_{\rm vg}^4\gamma_0{\cal G}_0 (0)|\omega+\Omega|}}, 
\end{eqnarray}
where $c_y \simeq 1/2$, 
\begin{eqnarray}
& & c_x \simeq \frac{\Delta_p}{\displaystyle{2\biggl(\Delta_p+\Delta_l\sqrt{\frac{2\pi}{h}}\biggr)}}, 
\end{eqnarray}
and 
\begin{eqnarray}
r_g &=& \xi_{\rm vg}^{-2} = 1-\frac{({\cal G}_0 (0))^2}{2\pi r_{\small H}^2 s}\biggl(\Delta_p+\Delta_l\sqrt{\frac{2\pi}{h}}\biggr). 
\end{eqnarray}

As shown in Ref.10, this glass fluctuation in 3D systems is 2D-like if the point disorder is absent. Reflecting this feature, the coefficient $c_x$ vanishes in $\Delta_p$, $\Delta_p^{(\Phi)} \to 0$ limit.

Diagrams contributing to the r.h.s. of eq.(20) are described in Fig.2, where each open rectangle corresponds to ${\tilde{\cal G}}_{\rm vg}$. Eight pairs of diagrams shown in Fig.2 take the form 
\begin{eqnarray}
1)+2) &\simeq& \int_{\bf k}\int_{\bf k^{\prime}}\frac{k_+ k_-^{\prime}}{2}{\mit V}({\bf k}){\mit V}_{\Phi}({\bf k^{\prime}})\biggl(-\frac{\partial}{\partial|\Omega|}\biggr)\beta^{-1}\sum_{\omega}({\cal G}_0(\omega))^2{\tilde{\cal G}}_{\rm vg}({\bf k}-{\bf k^{\prime}};\omega,\omega+\Omega), \nonumber \\
1)^{\prime}+2)^{\prime} &\simeq& \int_{\bf k}\int_{\bf k^{\prime}}\frac{k_+ k_+^{\prime}}{2}{\mit V}({\bf k}){\mit V}_{\Phi}({\bf k^{\prime}})\biggl(-\frac{\partial}{\partial|\Omega|}\biggr)\beta^{-1}\sum_{\omega}{\cal G}_0 (\omega){\cal G}_0 (\omega+\Omega){\tilde{\cal G}}_{\rm vg}({\bf k}-{\bf k^{\prime}};\omega,\omega+\Omega), \nonumber \\
3)+4) &\simeq& \int_{\bf k}\int_{\bf k^{\prime}}\frac{k_+ k_-^{\prime}}{2}{\mit V}({\bf k}){\mit V}_{\Phi}({\bf k^{\prime}})\biggl(-\frac{\partial}{\partial|\Omega|}\biggr)\beta^{-1}\sum_{\omega}({\cal G}_0 (\omega+\Omega))^2{\tilde{\cal G}}_{\rm vg}({\bf k}-{\bf k^{\prime}};\omega,\omega+\Omega), \nonumber \\
3)^{\prime}+4)^{\prime} &\simeq& \int_{\bf k}\int_{\bf k^{\prime}}\frac{k_+ k_+^{\prime}}{2}{\mit V}({\bf k}){\mit V}_{\Phi}({\bf k^{\prime}})\biggl(-\frac{\partial}{\partial|\Omega|}\biggr)\beta^{-1}\sum_{\omega}{\cal G}_0 (\omega){\cal G}_0 (\omega+\Omega){\tilde{\cal G}}_{\rm vg}({\bf k}-{\bf k^{\prime}};\omega,\omega+\Omega). \nonumber \\
3)^{\prime}+4)^{\prime} &\simeq& \int_{\bf k}\int_{\bf k^{\prime}}\frac{k_+ k_+^{\prime}}{2}{\mit V}({\bf k}){\mit V}_{\Phi}({\bf k^{\prime}})\biggl(-\frac{\partial}{\partial|\Omega|}\biggr)\beta^{-1}\sum_{\omega}{\cal G}_0 (\omega){\cal G}_0 (\omega+\Omega){\tilde{\cal G}}_{\rm vg}({\bf k}-{\bf k^{\prime}};\omega,\omega+\Omega). \nonumber \\
5)+6) &\simeq& \int_{\bf k}\int_{\bf k^{\prime}}\frac{k_+ k_-^{\prime}}{2}{\mit V}({\bf k}){\mit V}_{\Phi}({\bf k^{\prime}})\biggl(-\frac{\partial}{\partial|\Omega|}\biggr)\beta^{-1}\sum_{\omega}{\cal G}_0(\omega){\cal G}_0(\omega+\Omega){\tilde{\cal G}}_{\rm vg}({\bf k}-{\bf k^{\prime}};\omega,\omega+\Omega), \nonumber \\
5)^{\prime}+6)^{\prime} &\simeq& \int_{\bf k}\int_{\bf k^{\prime}}\frac{k_+ k_+^{\prime}}{2}{\mit V}({\bf k}){\mit V}_{\Phi}({\bf k^{\prime}})\biggl(-\frac{\partial}{\partial|\Omega|}\biggr)\beta^{-1}\sum_{\omega}({\cal G}_0 (\omega))^2{\tilde{\cal G}}_{\rm vg}({\bf k}-{\bf k^{\prime}};\omega,\omega+\Omega), \nonumber \\
7)+8) &\simeq& \int_{\bf k}\int_{\bf k^{\prime}}\frac{k_+ k_-^{\prime}}{2}{\mit V}({\bf k}){\mit V}_{\Phi}({\bf k^{\prime}})\biggl(-\frac{\partial}{\partial|\Omega|}\biggr)\beta^{-1}\sum_{\omega}{\cal G}_0 (\omega){\cal G}_0 (\omega+\Omega){\tilde{\cal G}}_{\rm vg}({\bf k}-{\bf k^{\prime}};\omega,\omega+\Omega), \nonumber \\
7)^{\prime}+8)^{\prime} &\simeq& \int_{\bf k}\int_{\bf k^{\prime}}\frac{k_+ k_+^{\prime}}{2}{\mit V}({\bf k}){\mit V}_{\Phi}({\bf k^{\prime}})\biggl(-\frac{\partial}{\partial|\Omega|}\biggr)\beta^{-1}\sum_{\omega}({\cal G}_0 (\omega+\Omega))^2{\tilde{\cal G}}_{\rm vg}({\bf k}-{\bf k^{\prime}};\omega,\omega+\Omega), \nonumber 
\end{eqnarray}
where we have used the fact that the wave number ${\bf k}-{\bf k^{\prime}}$ carried by the glass fluctuation is small so that 
\begin{eqnarray}
{\tilde \Delta}_1 ({\bf k}){\tilde \Delta}_0 ({\bf k^{\prime}})-{\tilde \Delta}_{01}({\bf k}){\tilde \Delta}_{01}({\bf k^{\prime}}) \simeq 4 h^2 {\mit V}({\bf k}){\mit V}_{\Phi}({\bf k^{\prime}}),
\end{eqnarray}
and 
\begin{eqnarray}
{\mit V}({\bf k}) &=& \Delta_p \exp\biggl(-\frac{k^2}{2}\biggr)+\frac{2\pi}{\sqrt{h}}\Delta_l\exp\biggl(-\frac{k_x^2}{2}\biggr)\delta(k_y), \\
{\mit V}_{\Phi}({\bf k}) &=& \Delta_p^{(\Phi)}\exp\biggl(-\frac{k^2}{2}\biggr)+\frac{2\pi}{\sqrt{h}}\Delta_l^{(\Phi)}\exp\biggl(-\frac{k_x^2}{2}\biggr)\delta(k_y). 
\end{eqnarray}
Further, the frequency dependence of the high energy fluctuation in the next lowest LL was neglected so that ${\cal G}_1 = \langle|\varphi_1|^2 \rangle$ is the constant $\mu_1^{-1}$Cbecause we are interested in the time scales of the glass fluctuation with the lowest energy in the present situation.

Below, we focus on the real part of the conductivities. Then, $R_Q\,\sigma_{{\rm vg},\,xx}$ is given by 
\begin{eqnarray}
& & \hspace{10mm} 1)+\cdots+8)-\bigl(\ 1)^{\prime}+\cdots+8)^{\prime}\ \bigr) \nonumber \\
&=& h^2 \int_{\bf k}\int_{\bf k^{\prime}}k_y k_y^{\prime}\,{\mit V}({\bf k}){\mit V}_{\Phi}({\bf k^{\prime}})\biggl(-\frac{\partial}{\partial|\Omega|}\biggr)\beta^{-1}\sum_{\omega}({\cal G}_0 (\omega)+{\cal G}_0 (\omega+\Omega))^2{\tilde{\cal G}}_{\rm vg}({\bf k}-{\bf k^{\prime}};\omega,\omega+\Omega) \nonumber \\ 
&\simeq& h^2 \int_{\bf k} \Delta_p \Delta_p^{(\Phi)}k_y^2\,e^{-k^2}\biggl(-\frac{\partial}{\partial|\Omega|}\biggr)\beta^{-1}\sum_{\omega}({\cal G}_0(\omega)+{\cal G}_0 (\omega+\Omega))^2 \int_{\bf k^{\prime}}{\tilde{\cal G}}_{\rm vg}({\bf k^{\prime}};\omega,\omega+\Omega), 
\end{eqnarray}
while the corresponding expression of $R_Q\, \sigma_{{\rm vg},\,yy}$ is 
\begin{eqnarray}
& & \hspace{10mm} 1)+\cdots+8)+\bigl(\ 1)^{\prime}+\cdots+8)^{\prime}\ \bigr) \nonumber \\
&=& h^2 \int_{\bf k}\int_{\bf k^{\prime}}k_x k_x^{\prime}\,{\mit V}({\bf k}){\mit V}_{\Phi}({\bf k^{\prime}})\biggl(-\frac{\partial}{\partial|\Omega|}\biggr)\beta^{-1}\sum_{\omega}({\cal G}_0 (\omega)+{\cal G}_0 (\omega+\Omega))^2{\tilde{\cal G}}_{\rm vg}({\bf k}-{\bf k^{\prime}};\omega,\omega+\Omega) \nonumber \\
&\simeq& h^2 \int_{\bf k} \Delta_p \Delta_p^{(\Phi)}k_x^2\,e^{-k^2}\biggl(-\frac{\partial}{\partial|\Omega|}\biggr)\beta^{-1}\sum_{\omega}({\cal G}_0 (\omega)+{\cal G}_0 (\omega+\Omega))^2 \int_{\bf k^{\prime}}{\tilde{\cal G}}_{\rm vg}({\bf k^{\prime}};\omega,\omega+\Omega) \nonumber \\
&+& h^{3/2} \int_{\bf k} (\Delta_{p}\Delta_l^{(\Phi)}+\Delta_{l}\Delta_p^{(\Phi)})k_x^2\,e^{-k_x^2}\biggl(-\frac{\partial}{\partial|\Omega|}\biggr)\beta^{-1}\sum_{\omega}({\cal G}_0(\omega)+{\cal G}_0 (\omega+\Omega))^2 \nonumber \\
&\times& \int_{\bf k^{\prime}}\,e^{-{k_y^{\prime}}^2 /2}\,{\tilde{\cal G}}_{\rm vg}({\bf k^{\prime}};\omega,\omega+\Omega) \nonumber \\
&+& h \int_{k_x} \Delta_l \Delta_l^{(\Phi)}k_x^2\,e^{-k_x^2}\biggl(-\frac{\partial}{\partial|\Omega|}\biggr)\beta^{-1}\sum_{\omega}({\cal G}_0(\omega)+{\cal G}_0(\omega+\Omega))^2\int_{k_x^{\prime}} {\tilde{\cal G}}_{\rm vg}(k_x^{\prime};\omega,\omega+\Omega).
\end{eqnarray}

\section{Numerical analysis of resistivity curves}

To discuss the resistive behaviors in the quantum regime in details, we examine the resulting resistivity curves numerically. Before performing this, the expressions of conductivities need to be arranged in a useful form.

First, strictly speaking, the pinning strengths such as $\Delta_p$ and $\Delta_p^{(\Phi)}$ carry the ${\bf k}$ or ${\bf k^{\prime}}$ dependences. However, they are not obtained in a closed form. In particular, it is not easy to know the momentum dependence of $\Delta_l$ and $\Delta_l^{(\Phi)}$. For these reasons, we write, e.g., the coefficient $\int_{\bf k} h^2 \Delta_p \Delta_p^{(\Phi)} k_y^2\exp(-k^2)$ in eq.(27) as ${\tilde c}_{\rm p} (\Delta_p)^2/(r_{\small H}^4 s^2)$.  Note that $\Delta_p/(\xi_0^2 s)$ is a dimensionless quantity. Since we expect the magnitude of $\Delta_p^{(\Phi)}$ to be much smaller than that of $\Delta_p$, the coefficient ${\tilde c}_{\rm p}$ is assumed to be smaller than unity. NextCarranging the frequecy summation using the replacement $\sum_{a < \omega < b} A(\omega) \to \sum_{\omega > a} A(\omega) -\sum_{\omega \ge b} A(\omega)$ and performing the ${\bf k^{\prime}}$-integral over the glass correlation function, the expression of $\sigma_{{\rm vg},\,xx}$ becomes 
\begin{eqnarray}
s\,R_Q\,\sigma_{{\rm vg},\,xx} &=& \frac{(\Delta_p)^2}{2\pi r_{\small H}^4 s^{2}}\biggl(1+\frac{\Delta_l}{\Delta_p}\sqrt{\frac{2\pi}{h}}\biggr)^{1/2}{\tilde c}_{\rm p}\beta^{-1}({\cal G}_0 (0))^2 \nonumber \\ 
&\times& \sum_{\omega}\frac{\partial}{\partial|\omega|}\biggl[\frac{{\cal G}_0 (\omega)({\cal G}_0 (\omega)+{\cal G}_0 (0))}{2}\ln\biggl(1+k_c^2(c_x c_y)^{1/2}\xi_{\rm vg}^2\biggr) \nonumber \\ 
&-& ({\cal G}_0 (\omega))^2\ln\biggl(\frac{1+k_c^2 (c_x c_y)^{1/2}\xi_{\rm vg}^2+2|\omega|\gamma_0{\cal G}_0 (0)\xi_{\rm vg}^4}{1+2|\omega|\gamma_0{\cal G}_0 (0)\xi_{\rm vg}^4}\biggr)\biggr],
\end{eqnarray}
where $k_c$ is a cut-off of order unity of the $|{\bf k}|$-integral. Further, the prefactors $\int_{k_x} h^{3/2} (\Delta_p \Delta_l^{(\Phi)} + \Delta_l \Delta_p^{(\Phi)}) k_x^2 \exp(-k_x^2)$ and $\int_{k_x} h \Delta_l \Delta_l^{(\Phi)} k_x^2 \exp(-k_x^2)$ will also be replaced with $2 h^{-1/2} \Delta_p \Delta_l {\tilde c}_{\rm lp}/(r_{\small H}^4 s^2)$ and $h^{-1} (\Delta_l)^2 {\tilde c}_{\rm l}/(r_{\small H}^4 s^2)$, respectively, by introducing unknown constants ${\tilde c}_{\rm lp}$ and ${\tilde c}_{\rm l}$ measuring nonlocalities of the pinning potentials due to line defects. Then, the conductivity $\sigma_{{\rm vg},\,yy}$ takes the form 
\begin{eqnarray}
s \, R_{Q} \, \sigma_{{\rm vg},\,yy} \! \! \! \! &=& \! \! \! \! s \, R_{Q} \, \sigma_{{\rm vg},\,xx}\biggl(1+2\frac{\Delta_l\,{\tilde c}_{\rm lp}}{\Delta_p \,{\tilde c}_{\rm p}}h^{-1/2}\biggr)+\xi_{\rm vg}\frac{{\tilde c}_{\rm l}h^{-1}(\Delta_l)^2}{\sqrt{2}r_{\small H}^4 s^{2}}\biggl(1+\frac{\Delta_l}{\Delta_p}\sqrt{\frac{2\pi}{h}}\biggr)^{1/2}\beta^{-1}({\cal G}_0 (0))^2 \nonumber \\
\! \! \! \! &\times& \! \! \! \! \sum_{\omega}\frac{\partial}{\partial|\omega|}\biggl[\frac{{\cal G}_0 (\omega)({\cal G}_0 (\omega)+{\cal G}_0 (0))}{2}-({\cal G}_0 (\omega))^2(1+2|\omega|\gamma_0{\cal G}_0 (0)\xi_{\rm vg}^4)^{-1/2}\biggr].
\end{eqnarray}
We guess that, compared with the point defects, real line defects will have a longer correlation range in the $x$ direction, because they, if existing at low densities, have a strong impact. Hence, we expect ${\tilde c}_{\rm l}$ to be somewhat larger compared with ${\tilde c}_{\rm p}$. When focusing on metallic superconducting films to be described in dirty limit, $\Delta_p$ is identified with $b_p$ defined in Ref.2, while the dimensionless strength $\Delta_l/(\xi_0^2 s)$ of the artificial line defects is unknown and will be assumed to be a constant independent of the magnetic field.

\begin{figure}
\scalebox{0.275}[0.275]{\includegraphics{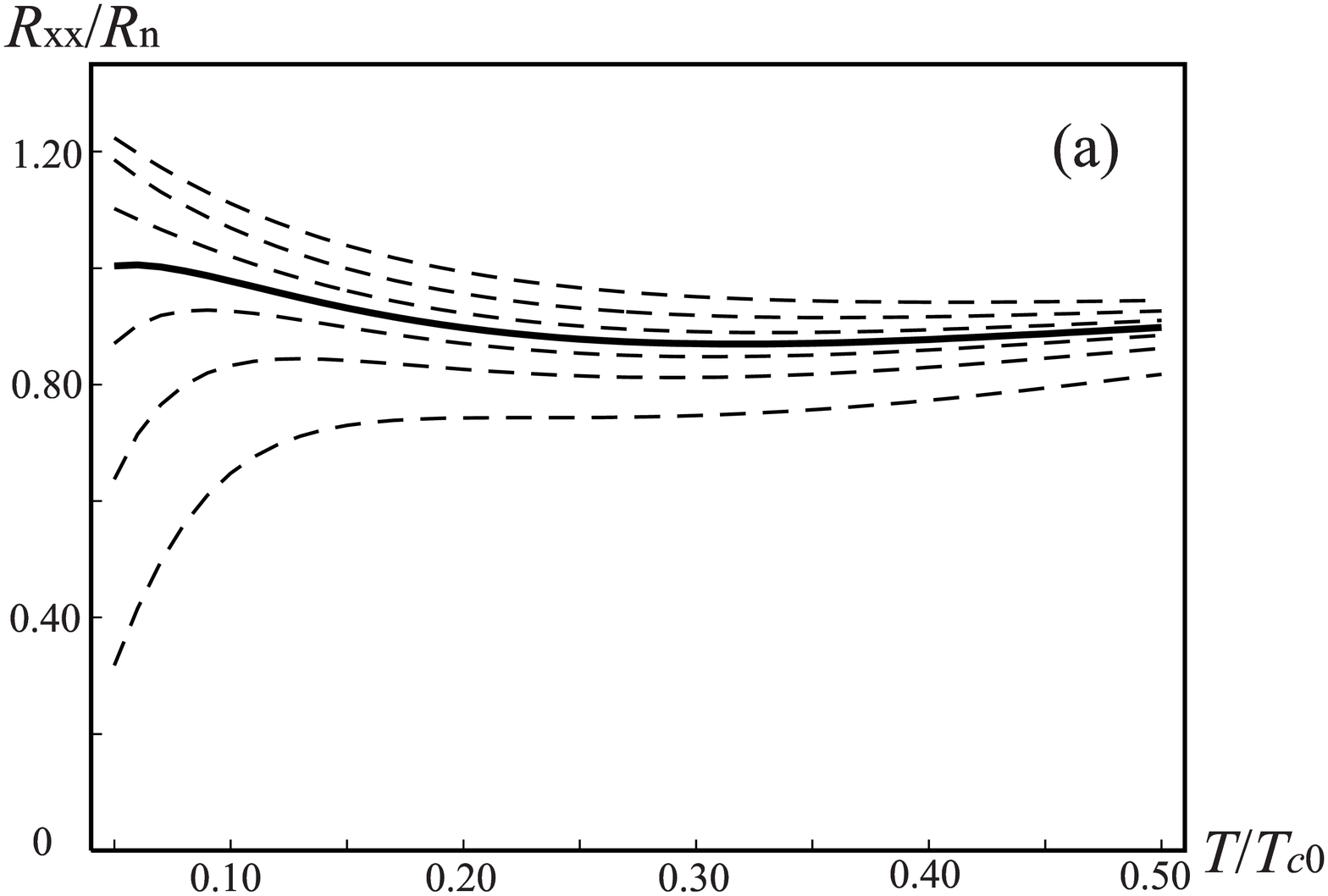}}
\hspace{10mm}
\scalebox{0.275}[0.275]{\includegraphics{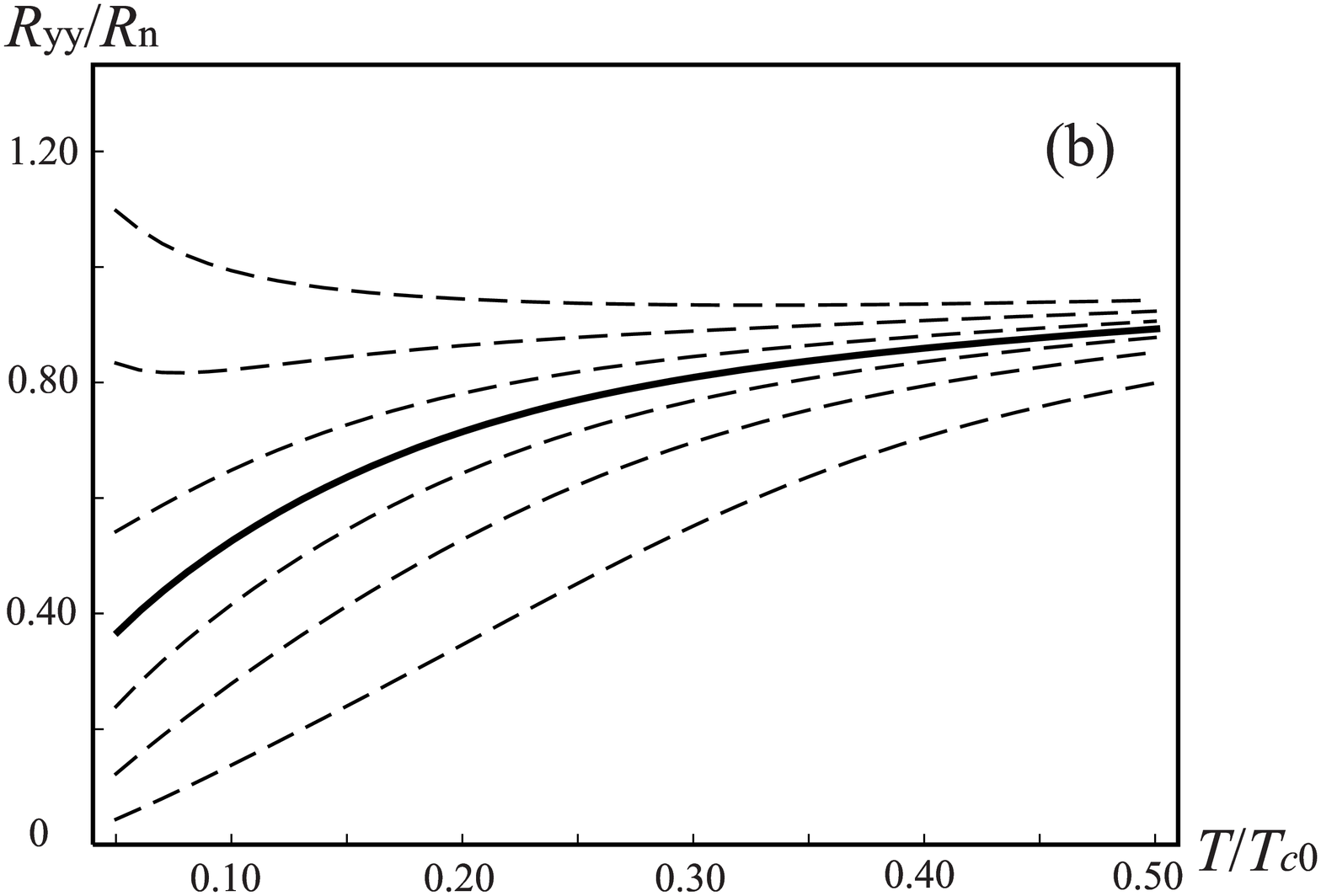}} \\
\scalebox{0.275}[0.275]{\includegraphics{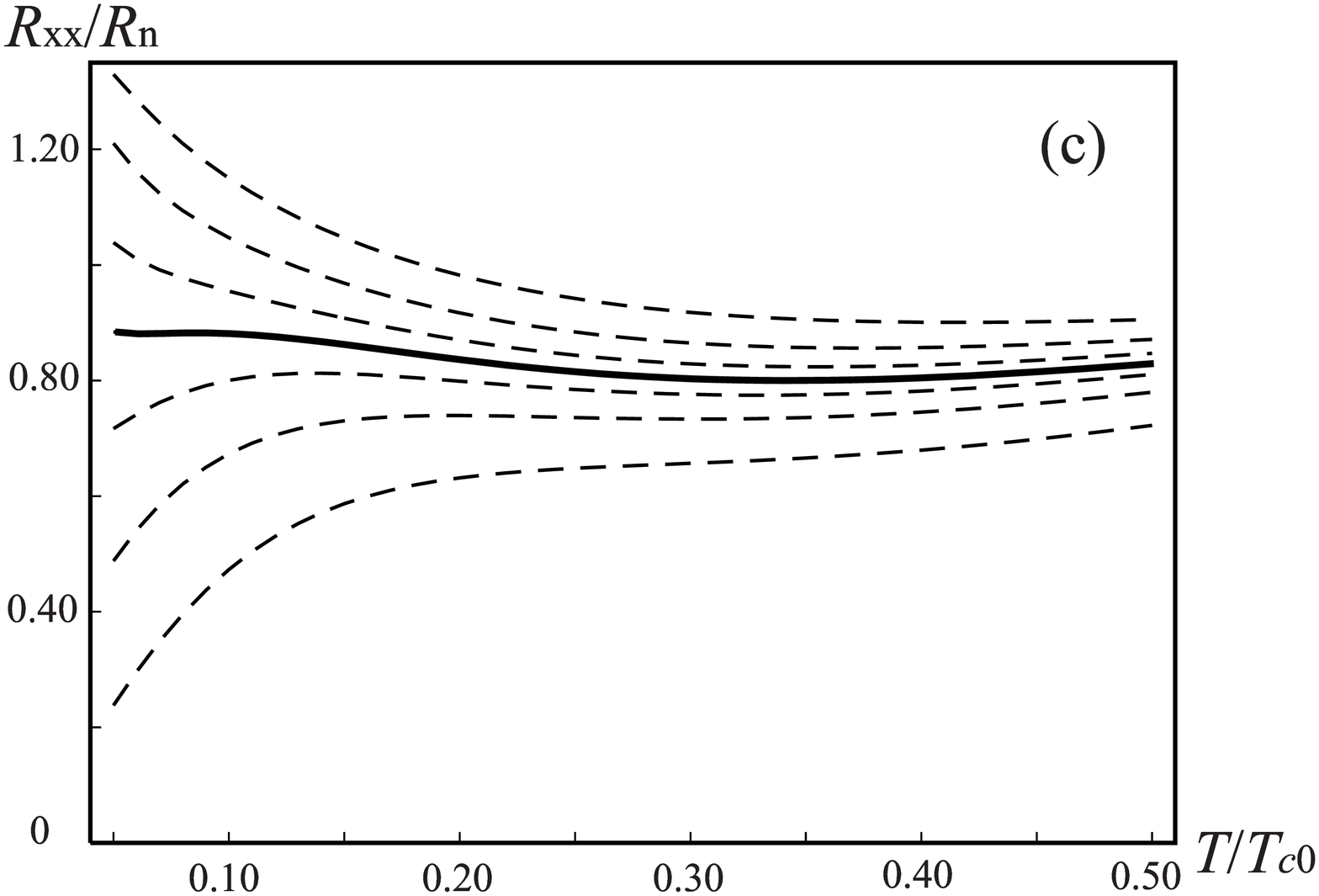}}
\hspace{10mm}
\scalebox{0.275}[0.275]{\includegraphics{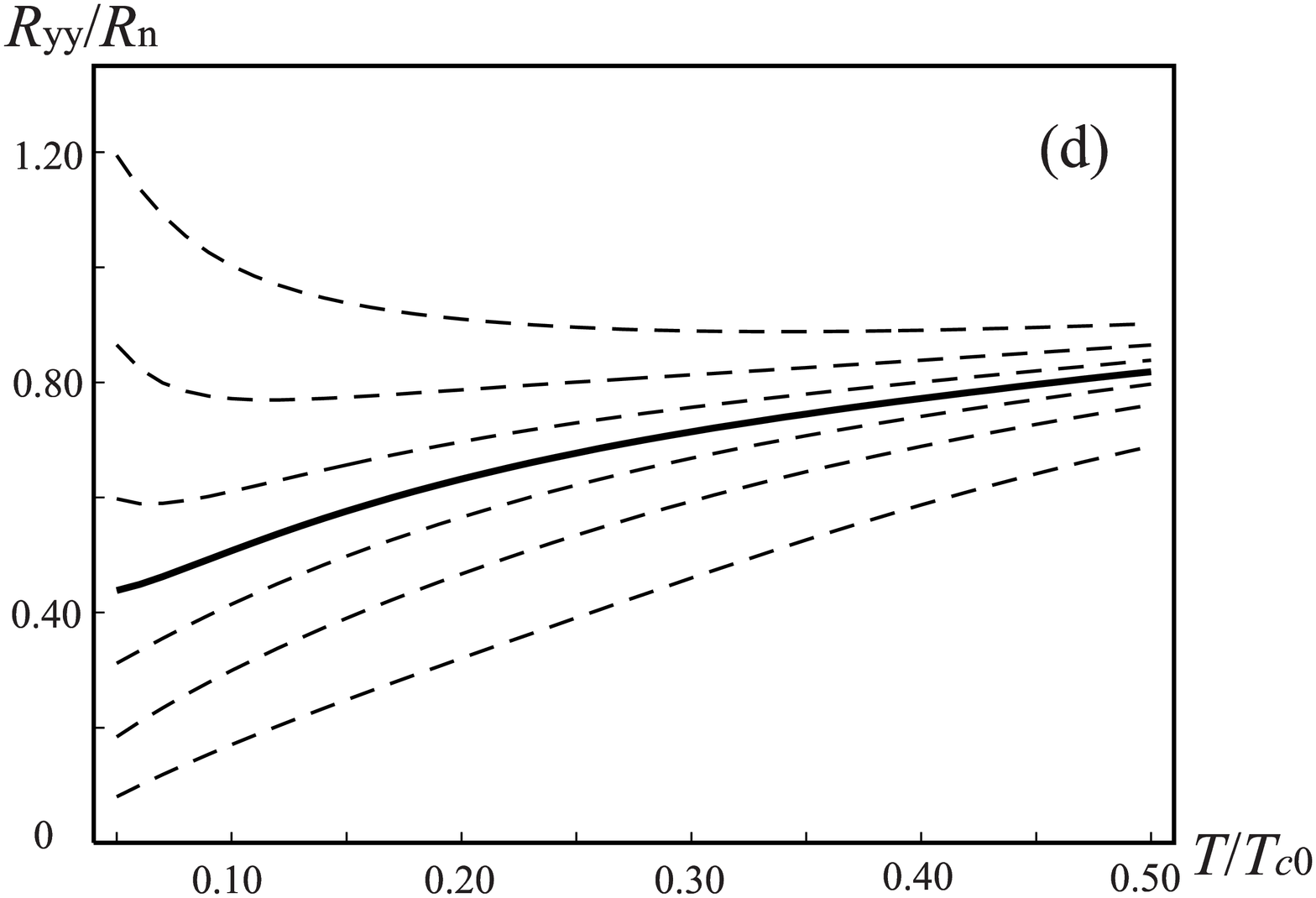}}
\caption{Examples of resistivity curves computed in terms of eqs.(29) and (30),  $R_{xx}$ ((a) and (c)) and $R_{yy}$ ((b) and (d)). The curves in (a) and (b) are those when $R_n = 0.39 R_Q$, and the values of the reduced field $h$ are $h=0.69$ (bottom), $0.74$, $0.77$, $0.79$, $0.81$, $0.84$, $0.88$ (top), while the curves in (c) and (d) are those when $R_n = 0.52 R_Q$, and the $h$ values $0.60$ (bottom), $0.65$, $0.68$, $0.70$, $0.72$, $0.75$, $0.80$ (top) were used. Each curve at $H=H_c$ is indicated by a solid curve in each figure. Other parameters used here are $\Delta_l/(\xi_0^2 s) = 3.0\times 10^{-3}$, ${\tilde c}_{\rm p} = 0.2$, ${\tilde c}_{\rm lp} = 0.2$ and ${\tilde c}_{\rm l} = 5.0$.
}
\end{figure}

Besides these expressions of the glass fluctuation terms of conductivities, we have included the contributions insensitive to the defects to the total conductivities, the quasiparticle term $\sigma_n$ and the ordinary SC fluctuation term $\sigma_{\rm fl}$ both of which were defined in Ref.12. In eq.(10) giving the SC fluctuation propagator ${\cal G}(\omega)$ selfconsistently, a fluctuation renormalization of the bare pinning strengths $\Delta_w$ ($w=p$, $l$) will be incorporated by replacing them by $\Delta_w/(1+v_p)$, where 
\begin{equation}
v_p = \frac{b \beta^{-1}}{2\pi r_{\small H}^2 s} \sum_{\omega}({\cal G}_0(|\omega|))^2. 
\end{equation}
Consistently with this treatment, the glass correlation length is give by 
\begin{eqnarray}
\xi_{\rm vg}^{-2} = 1-\frac{({\cal G}_0 (0))^2}{2\pi r_{\small H}^2 s(1+v_p)} \biggl(\Delta_p + \Delta_l \sqrt{\frac{2\pi}{h}}\biggr)+\sqrt{\gamma_0 T{\cal G}_0 (0)},
\end{eqnarray}
where the second ($\sim T^{1/2}$) term, which should appear in the vicinity of the quantum critical point of the glass transition, 
was incorporated \cite{IR}.

Using the above expressions of each term of the conductivities, together with the expressions of $\mu(0)$, $b$, $\Delta_p /(r_{\small H}^2 s)$, and $\gamma_0$ in dirty limit \cite{RI}, we have numerically examined the resulting resistance v.s. temperature curves in order to clarify the significant roles of line defects. Below, $R_n$ denotes the resistance in the normal state of a superconducting film with point defects but no line defects. We choose the parameter values for the dimensionless strength of line defects $\Delta_l/(\xi_0^2 s)=1.0\times 10^{-3}$, $3.0\times 10^{-3}$, or $4.5\times 10^{-3}$, the momentum cut-off $k_c = 1.0$, $R_n/R_Q = 0.52$, $0.47$, or $0.39$, ${\tilde c}_{\rm p} = 0.2$, ${\tilde c}_{\rm lp} = 0.2$, and ${\tilde c}_{\rm l} = 5.0$. For the reason mentioned earlier, a much smaller value than ${\tilde c}_{\rm l}$ is assumed for ${\tilde c}_{\rm p}$.

Resistivity curves following from eqs.(29) and (30) are shown in Fig.3. As expected, the conductivity $\sigma_{xx}$ for a current perpendicular to the line defects includes no explicit effects of line defects, as expected from the mean field picture that the vortex flow motion in this case is unaffected by the presence of line defects. Actually, the resistivity curves in a narrow window of the magnetic field around an estimated critical field $H=H_c$ show a flat behavior insensitive to the temperature at low enough temperatures, although, as already seen elsewhere \cite{RI}, the flat behavior near $H_c$ is not visible in the case with low $R_n$ where the quantum fluctuation is weaker. We note that, for $R_n = 0.52 R_Q$, $H_c = 0.70 \phi_0/(2\pi\xi_0^2)$, while $H_c = 0.79 \phi_0/(2\pi\xi_0^2)$ for $R_n = 0.39 R_Q$. On the other hand, the response $\sigma_{yy}$ to a current parallel to the line defects has a term proportional to the square of the strength of line defects, and this term proportional to ${\tilde c}_{\rm l}$ is dominant near the critical point at $T = 0$ where $\xi_{\rm vg}$ diverges. In fact, in situations where the glass fluctuation plays a significant role, the glass fluctuation terms $\sigma_{{\rm vg},jj}$ of conductivities show the following behaviors 
\begin{eqnarray}
& & R_Q\,s\,\sigma_{{\rm vg},\,xx} \sim X(T\,\xi_{\rm vg}^{\,z_{\rm vg}}), \nonumber \\
& & R_Q\,s\,\sigma_{{\rm vg},\,yy} \sim \xi_{\rm vg}Y(T\,\xi_{\rm vg}^{\,z_{\rm vg}}),
\end{eqnarray}
where $z_{\rm vg}$ is the dynamical critical exponent of the glass fluctuation and, in the present mean field (i.e., Gaussian) approximation, is four. In particular, just on the critical field defined at low enough temperatures, i.e., $H=H_c$, we have the relation  $\xi_{\rm vg} \sim T^{-1/z_{\rm vg}}$ (see also eq.(32)) when $T \to 0$. Then, the critical value of $\sigma_{{\rm vg}, \, jj}$ becomes 
\begin{eqnarray}
& & R_Q\,\sigma_{{\rm vg},\,xx}(H=H_c) \sim {\rm ( nonuniversal ) constant}, \\
& & R_Q\,\sigma_{{\rm vg},\,yy}(H=H_c) \sim T^{-1/z_{\rm vg}}. 
\end{eqnarray}

This divergent $\sigma_{{\rm vg},\,yy}(H=H_c)$, indicative of a vanishing critical resistance, corresponds to that \cite{Cha} expected in the virtual 1D superconductor-insulator quantum transition where the critical conductance is infinite. As seen in the figures, this 1D behavior in $\sigma_{yy}$ is more visible for a high density of line defects and/or for a smaller $R_n/R_Q$ value (i.e., weaker quantum fluctuation). As the figures suggest, the critical field seems to be well estimated from an onset field of the saturated behavior of $R_{xx}(T)$ curves, while in the case with low $R_n/R_Q$ the critical field is not well estimated from the corresponding $R_{xx}(T)$ curves in which the saturated (flat) behavior is vague and not clearly seen. That is, when the flat behavior in a vanishingly narrow field range is clearly visible in $R_{xx}(T)$ curves, such a flat behavior should be a direct consequence of the quantum vortex-glass fluctuation, while the position of critical field in systems with low $R_n/R_Q$ is more clearly seen in $R_{yy}$ data by artificially introducing line defects in the $y$-direction.

\section{Discussion}

\begin{figure}
\scalebox{0.275}[0.275]{\includegraphics{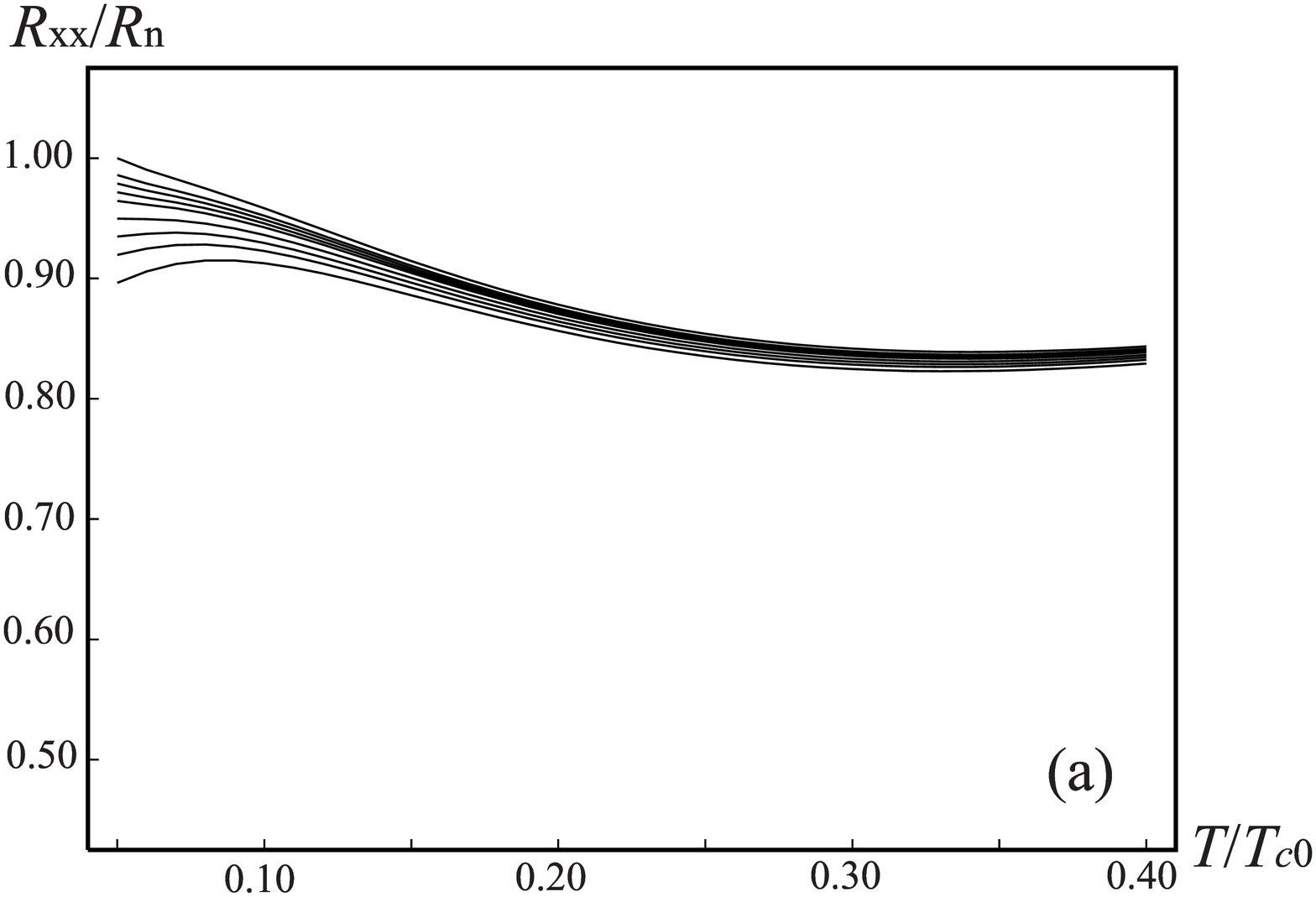}}
\scalebox{0.275}[0.275]{\includegraphics{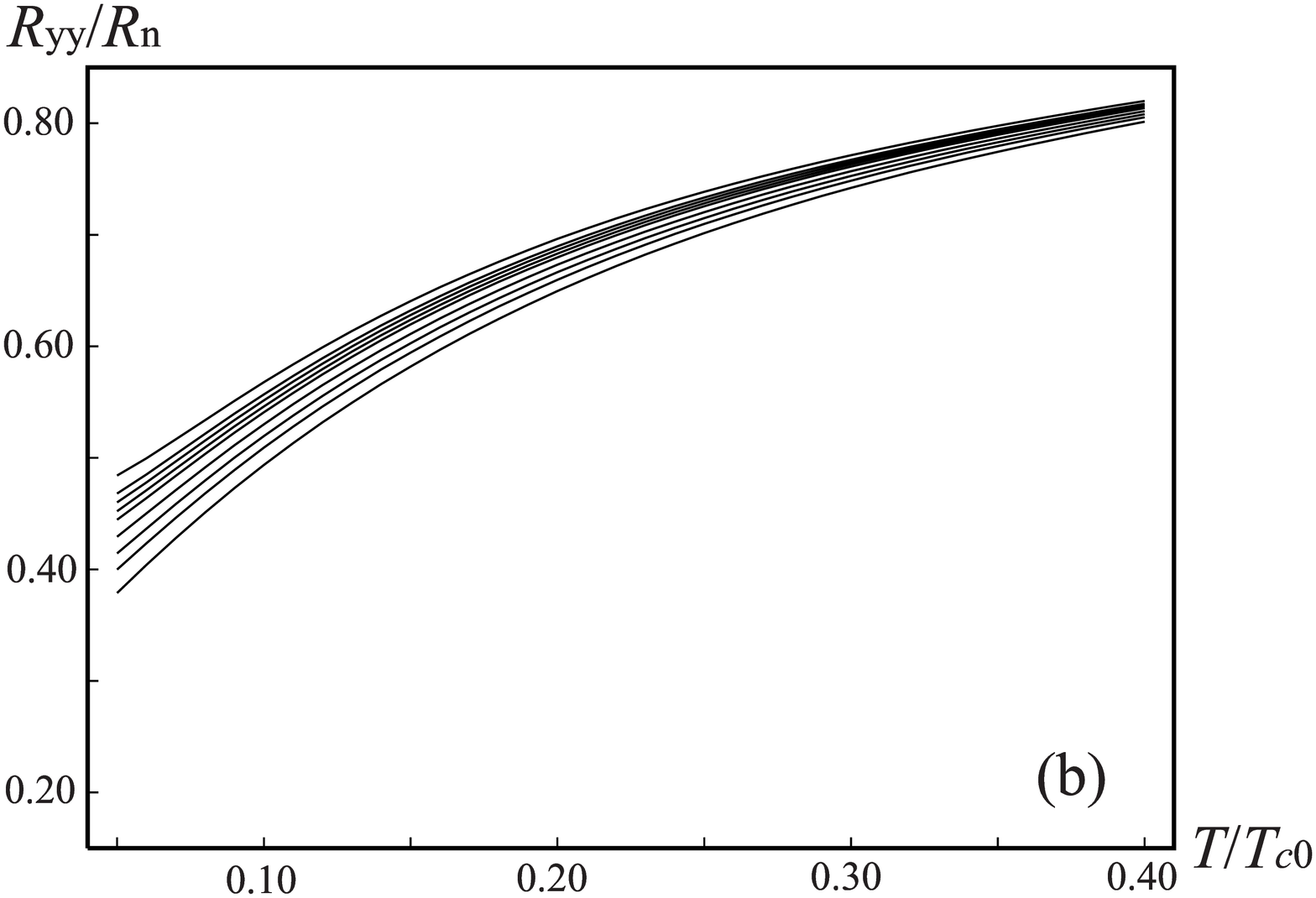}} \\
\scalebox{0.275}[0.275]{\includegraphics{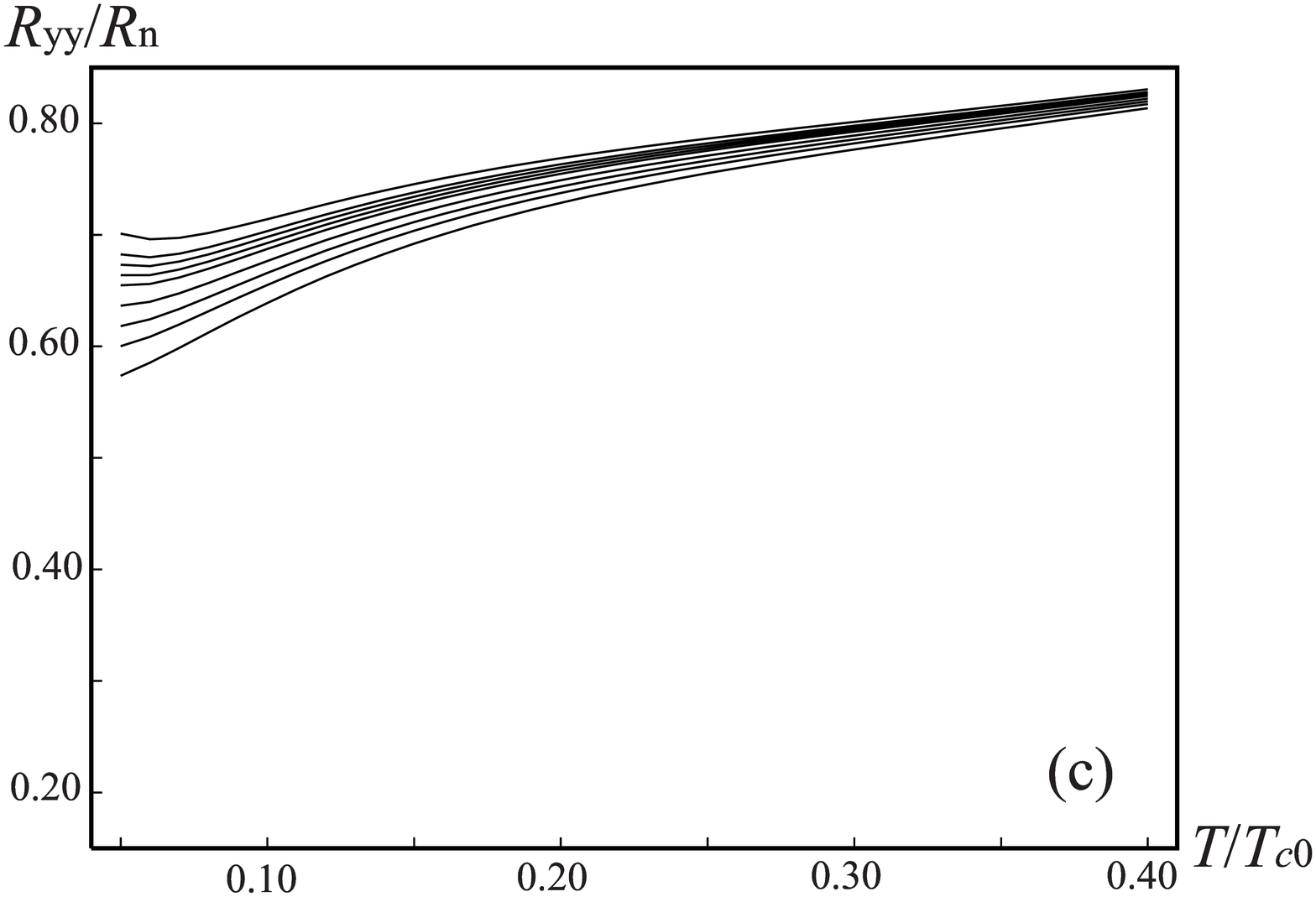}}
\caption{Computed (a) $R_{xx}$ curves, (b) $R_{yy}(T)$ curves in the case with persistent line defects parallel to ${\hat y}$, and (c) $R_{yy}(T)$ ones in the case with line-like defects with ${\tilde l}_y = 0.75$ (i.e., in the situation sketched in Fig.1 (b)) at the fields $h=0.73$, $0.733$, $0.735$, $0.737$, $0.739$, $0.74$, $0.741$, $0.742$, and $0.744$. The parameter values used here are $R_n = 0.47 R_Q$, $\Delta_l/(\xi_0^2 s) = 3.0\times 10^{-3}$, ${\tilde c}_{\rm p} = 0.2$, ${\tilde c}_{\rm lp} = 0.2$ and ${\tilde c}_{\rm l} = 5.0$.
}
\end{figure}

As an application of the present result, it is interesting to imagine the situation illustrated in Fig.1(b) where, instead of the line defects, line {\it like} defects with a long but finite correlation length $l_y$ of its orientation parallel to the $y$-axis are present in a SC film. As in the case with line defects parallel to the $y$-axis, there are no essential changes in $\sigma_{xx}$, and effects of the finite $l_y$ appear mostly in $\sigma_{{\rm vg}, \, yy}$. This finite $l_y$ is naturally incorporated by multiplying the last term of $w({\bf r}-{\bf r^{\prime}})$ ($w_\Phi({\bf r}-{\bf r^{\prime}})$), proportional to $\Delta_l$ ($\Delta_l^{(\Phi)}$), by $\exp(\,-(y-y^{\prime})^2/(2\,l_y^2) \,)$. Then, as far as $\Delta_l \sqrt{2 \pi/h} \ll \Delta_p$, effects of the finite $l_y$ on ${\cal G}_0(\omega)$ and $\xi_{\rm vg}$ are unimportant, and the essential change can be incorporated by replacing the last line of eq.(30) with 
\begin{eqnarray}
&\times& \frac{{\tilde l}_y}{\xi_{\rm vg}} \int\frac{dq}{\sqrt{2\pi}} \exp\biggl(-\frac{{\tilde l}_y^2\,q^2}{\xi_{\rm vg}^2}\biggr) \sum_\omega \frac{\partial}{\partial |\omega|} \biggl[ \frac{{\cal G}_0(\omega)({\cal G}_0 (\omega)+{\cal G}_0(0))}{2(1+q^2)^{1/2}} \nonumber \\ 
&-& \frac{({\cal G}_0(\omega))^2}{(1+q^2+2|\omega|\gamma_0{\cal G}_0(0) \xi_{\rm vg}^4)^{1/2}} \biggr],
\end{eqnarray}
where ${\tilde l}_y = l_y/r_{\small H}$.

From this expression, the following qualitative behaviors are expected on $R_{yy}$. At higher temperatures where $\xi_{\rm vg} \ll {\tilde l}_y$, the 1D behavior, eq.(33) or (35), of $\sigma_{yy}(H \simeq H_c)$ is valid, and the $R_{yy}(T)$ curve decreases upon cooling as in Fig.4(b) in fields around and below $H_c$. In contrast, at lower temperatures where $\xi_{\rm vg} \gg {\tilde l}_y$ close to or below $H_c$, the growth of $\sigma_{{\rm vg}, yy}$ due to the line-like defects upon cooling saturates in some field range around $H_c$, reflecting the ordinary 2D behavior. In fact, as the $R_{yy}$ data in Fig.4(c) following from eq.(30) with eq.(36) show, such resistivity curves around $H = H_c \simeq 0.74 \phi_0/(2\pi\xi_0^2)$ that are flat at the lowest temperatures but {\it decrease} upon cooling at higher temperatures become more remarkable when, as in Fig.1(b), line-like defects with a finite $l_y$ are included. This feature arising by assuming the presence of highly anisotropic defects may be relevant to the data \cite{Mason,Vicente} suggesting the presence of an intermediate ``bose-metal" phase.

Further, when the strength of line-like defects is stronger, reentrant resistivity curves, showing a decrease in intermediate temperatures but an {\it upturn} in lower temperatures, are more remarkable, just as in the curve at $h\geq 0.76$ in Fig.5(c), because stronger line defects may induce a sharp decrease of resistivity even in $H > H_c$ (i.e., $h > 0.746$). Such an observed reentrant resistivity curve, seen in earlier data of nominally amorphous materials,\cite{Hebard} is often regarded as implying that the sample is a granular film.\cite{Goldman} Although, roughly speaking, the present picture on the reentrant behavior is similar to assuming a granular structure in the film sample in the sense that a mesoscopic structure is invoked, we believe that the present picture is realistic in the sense that a local anisotropy of inhomogenuity is incorporated.

In addition, we note that, in the case corresponding to Fig.5(c), the critical resistance value, which is $\simeq 0.5 R_n$ in Fig.5(c), is highly suppressed by the presence of ${\tilde l}_y$ of order unity. As already pointed out in Ref.12, the apparent value of the critical sheet resistance is close to $R_n$ in most cases. However, some data show a significantly low critical resistance especially in systems with a larger $R_n/R_Q$ \cite{Wu}. The present result will give one of possible explanations of such a significantly low critical resistance estimated experimentally. 

\begin{figure}
\scalebox{0.275}[0.275]{\includegraphics{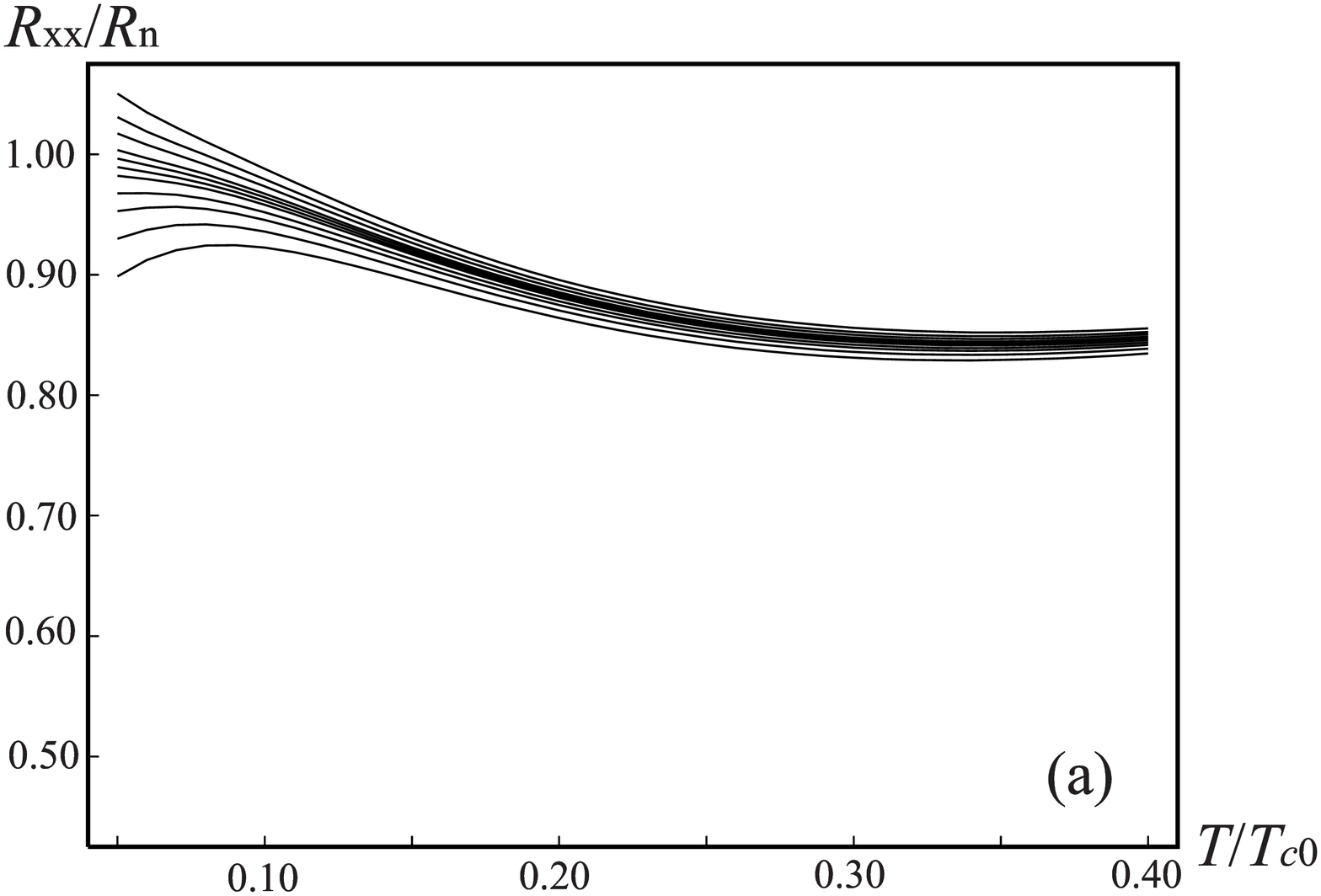}}
\scalebox{0.275}[0.275]{\includegraphics{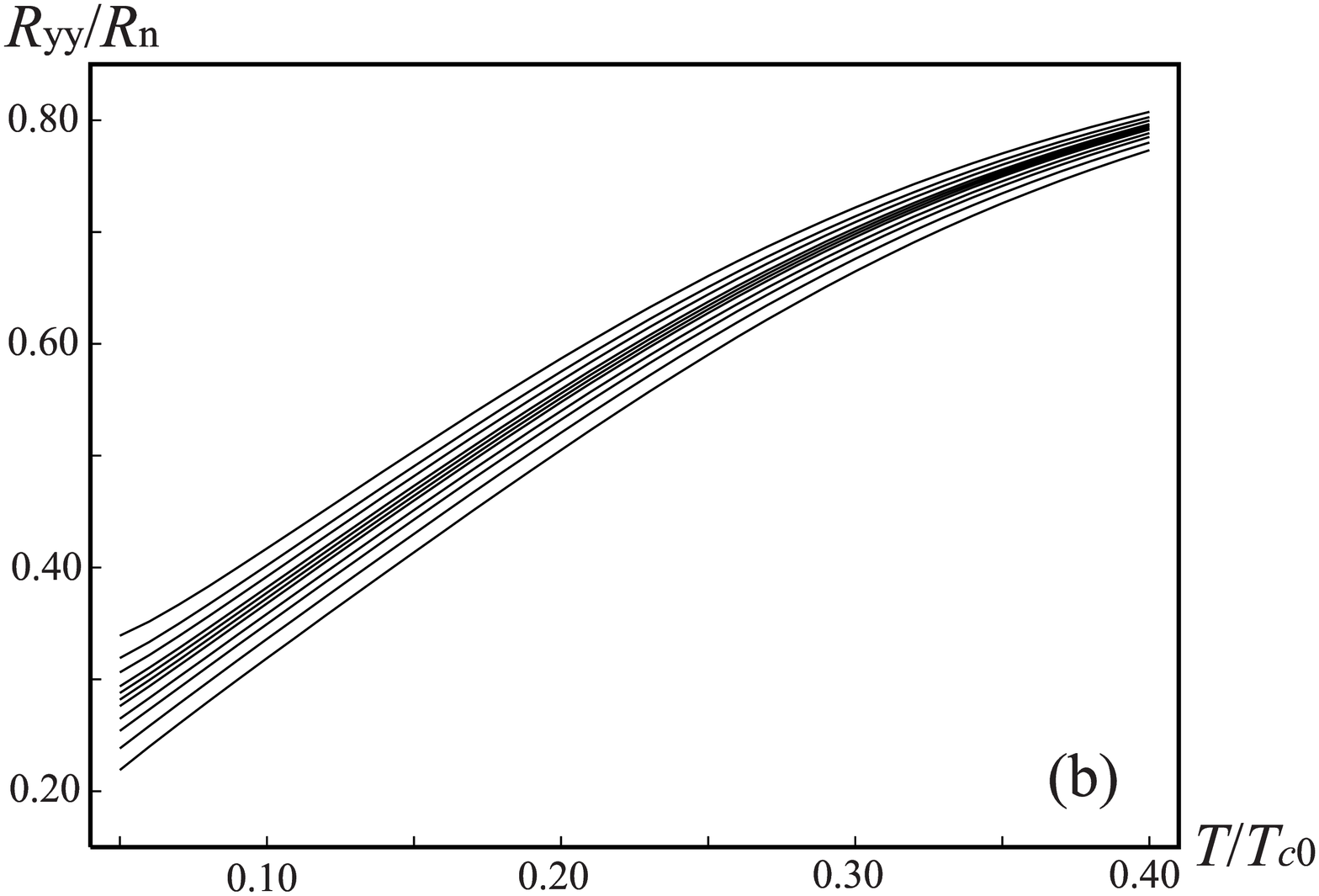}} \\
\scalebox{0.275}[0.275]{\includegraphics{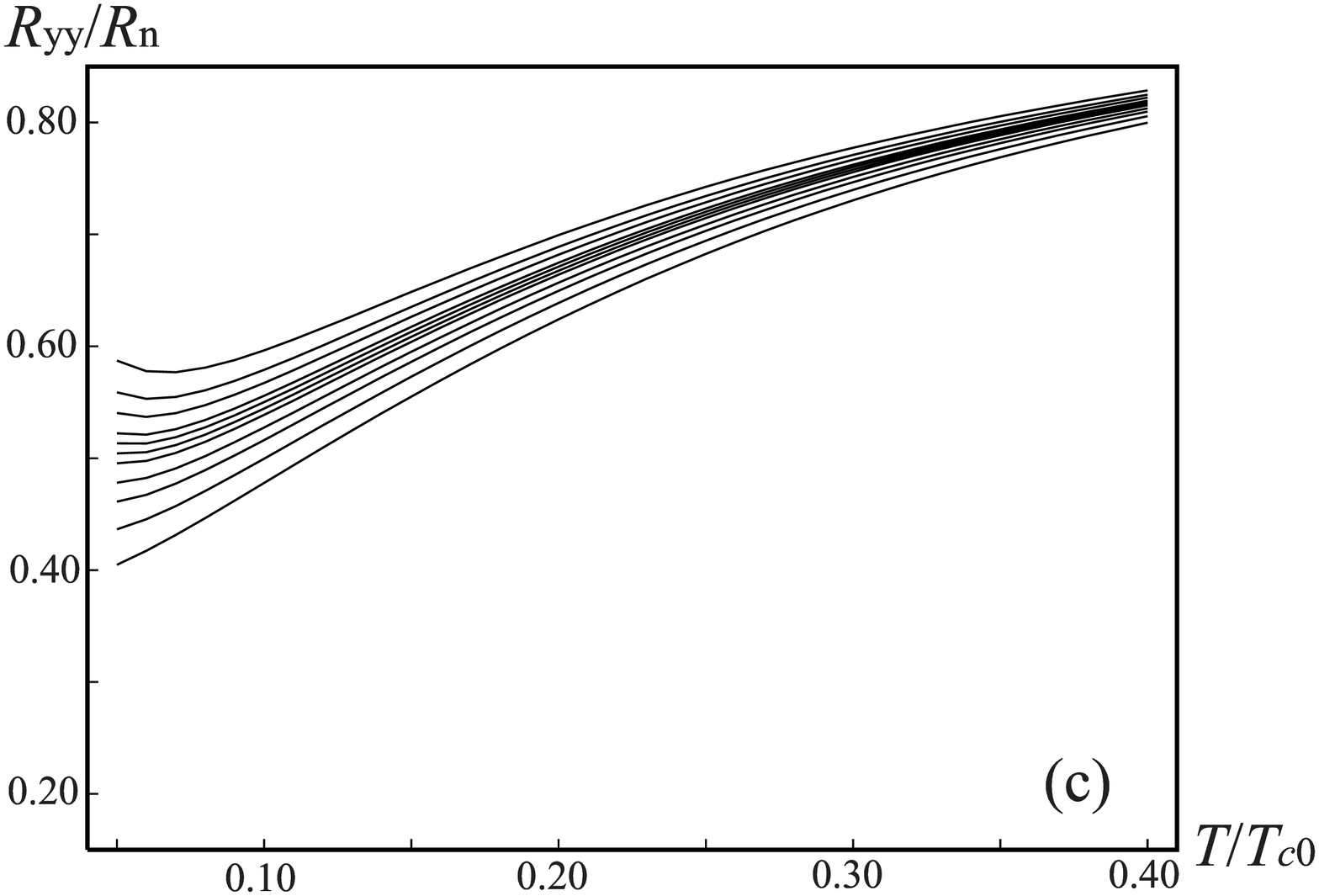}}
\caption{ Computed resistivity curves which are the same as those in Fig.4 except the use of  $\Delta_l/(\xi_0^2 s) = 4.5\times 10^{-3}$. $h = 0.737$, $0.741$, $0.744$, $0.746$, $0.748$, $0.749$, $0.75$, $0.751$, $0.753$, $0.755$ and $0.758$ respectively. 
}
\end{figure}

\begin{figure}
\scalebox{2.0}[2.0]{\includegraphics{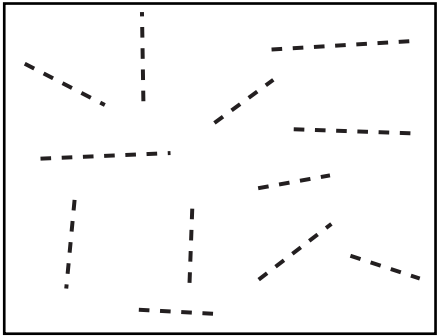}}
\caption{Sketch of a SC film including line-like defects with randomness in their directions. 
}
\end{figure}

Although an appearance of the ``bose-metal" behavior or a reentrant one has been demonstrated above for the case with line-like defects directed along a unique direction (see Fig.1(b)),  similar behaviors should also be seen in the case, as in Fig.6, where the correlated direction of line-like defects is random, because the regions with line-like defects parallel to the applied current have the largest weight on the total conductivity and thus, play dominant roles in the in-plane resistivity curves for any current direction. Therefore, if the film sample is globally isotropic but has such a highly anisotropic inhomogeneous structure over a finite length scale longer than the averaged vortex spacing, we argue that unexpected resistivity curves showing the ``bose-metal" behavior or a reentrant one may appear, reflecting the 1D to 2D dimensional crossover induced by the line-like defects correlated over such a finite length scale.

The computation in this work has been done using the facilities of the Supercomputer Center, Institute for Solid State Physics, University of Tokyo.

\end{document}